\documentclass[twoside]{dis09}
\usepackage[latin1]{inputenc}
\usepackage[dvips]{graphicx,epsfig,color}
\usepackage{wrapfig,rotating}
\usepackage{amssymb,amsmath,array}

\begin{document}
\title{An Electron-Ion Collider at Jefferson lab}

\author{Anthony W.~Thomas
%
%
\vspace{.3cm}\\
%
Suite 1, Jefferson Lab, 12000 Jefferson Ave., Newport News VA 23606 USA \\
and College of William and Mary, Williamsburg VA 23187 USA
%
}

\maketitle

\begin{abstract}
Long term plans for the investigation of the quark and gluon structure of matter 
have for some time focussed on the possibility of an electron-ion collider, with the 
nuclear physics communities associated with JLab and BNL being particularly 
active. We briefly outline the current thinking on this subject at Jefferson lab.
\end{abstract}

\section{Introduction}

%
%
%
%

As explained in the oral presentation~\cite{url}, 
the original plans for an electron-ion collider (ELIC) 
at Jefferson Lab involved the 
construction of two figure-of-eight rings, intersecting at up to four collision 
points, with a proton energy of 30-225 GeV (30-100 GeV/A for ions up to Pb) 
and electrons (and positrons) from 3 to 9 GeV~\cite{Bogacz:2007zza}. 
Since the construction 
of such a facility must await completion of the 12 GeV Upgrade 
at Jefferson Lab~\cite{CDR}, as 
well as the construction of FRIB, 
it is unlikely to begin much before the end of the 
next decade. The design for ELIC was appropriately 
ambitious for a world-leading 
machine that will not take data until the third decade of this millenium, with
a design luminosity approaching $10^{35}$ cm$^{-2}$ sec$^{-1}$. The exploration 
of the physics case for such a machine, 
which shares at least some common ground 
with the proposed eRHIC machine at BNL, involves an on-going collaboration 
under the heading of EIC~\cite{EIC}, between the communities associated 
with both national laboratories.

Since the cost of such very high energy colliders is likely to be rather high,  
at Jefferson Lab considerable effort has recently gone into the design of  
possible staging options, which present a strong, self-contained physics case,
yet have a cost comparable to that of FRIB. 
A multitude of novel suggestions for 
such a machine led to a very vigorous discussion between physicists, engineers 
and machine designers and the full user community at 
Jefferson Lab is just beginning 
to participate in these discussions. 
Nevertheless, there is already a very exciting design 
which currently appears to optimize the opportunities for scientific discovery 
while satisfying reasonable cost constraints. 
Everything which I shall present is 
the result of an impressive team effort by the group of people thanked in the 
acknowledgements.  
This new machine is currently known as the MEIC, or 
the medium energy electron-ion 
collider.

\section{MEIC Design Parameters}
As illustrated in Fig.~1, 
MEIC is designed to serve as the natural first stage of full 
ELIC construction. As far as possible the tunnel and components can be re-used 
in the ultimate machine, should that be built. The circumference of the machine is 
634m, with straight sections of 150m. The proton energies range from 12 to 60 GeV, 
in collision with electrons (or positrons) 
taken from the CEBAF 12 GeV Upgrade between 3 
and 11 GeV. This allows for e-p collisions over an impressive range of cm energies, 
from $s \, = \, 100 \, {\rm to} \,  2640$ GeV$^2$.  
We also note that if it were to be essential, 
even the MEIC can be staged with a warm ion ring, allowing proton 
momenta up  to 12 GeV/c, being a somewhat less expensive first step.
\begin{figure}
\centerline{\includegraphics{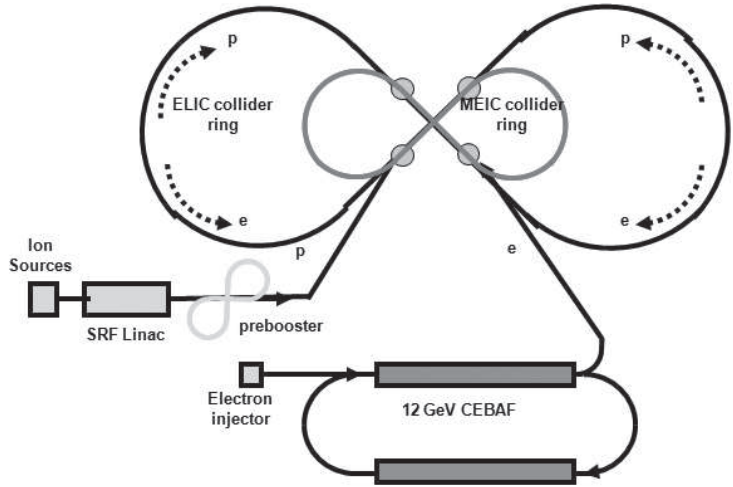}}
\caption{Layout of the ELectron Ion Collider (ELIC) and the MEIC 
at Jefferson Lab. Both of these machines would use polarized electrons 
(or positrons) at energies up to 11 GeV from the upgrade of CEBAF 
which is currently underway.}
\label{Fig:MEIC}
\end{figure}

As outlined earlier, a key design criterion has been to ensure that the luminosity 
will be appropriate for a world-leading macine in the 2020's and 30's. The current 
design yields excellent luminosity, around $10^{35}$ cm$^{-2}$-sec${-1}$, for 
$s \in (200,1200)$ GeV$^2$. The luminosity remains above 10$^{33}$ over 
the rest of the range. In terms of physics reach, this means that one can access 
the structure of polarized protons at luminosities of order $10^{35}$ above 
$x = 0.0008 \, {\rm at} \, Q^2 = 1$ GeV$^2$ and above 
$x = 0.01 \, {\rm at} \, 12$ GeV$^2$. For the time being, 
the luminosity is bounded by detector and data acquisition 
limitations, with the repetition rate being kept at a conservative 500 MHz. There 
is an urgent need for R \& D to explore whether one could raise this rate and 
consequently the luminosity.  

The figure of 8 structure, illustrated in Fig.~1, is designed to ensure 
high polarization for both the light ion and electron beams. There are 
four possible intersection points. It is intended to also make a polarized positron 
capability with the same luminosity available. 
 
A particularly attractive feature of the MEIC design
is that unlike ELIC or eRHIC, 
where some of the technical issues presently seem to 
be ``very challenging'' or worse, there is no 
issue associated with MEIC which 
ranks above ``challenging'' -- 
with the electron cooling and travelling focussing 
being the two major issues. Innovative features, such as crab crossing and 
crab cavities seem to be almost in hand. 
Other important issues which need work 
in the near future, include beam-beam effects 
and the formation of the high intensity 
low energy ion beam but these are not regarded as especially challenging.

\section{Physics at the MEIC}
As the novel design of the MEIC is very recent, the detailed physics case matched 
to its unique capabilities still needs considerable work and we certainly invite all 
interested members of the community to join this effort. 

In many ways the proposed MEIC is the perfect complement to the 12 GeV Upgrade 
currently underway at Jefferson Lab~\cite{Thomas:2007zza}. 
While the latter aims to define the spin 
and flavor dependence of nucleon and nuclear parton distribution functions in 
the valence region, 
the former is ideally suited to serve the same function for 
the sea. Indeed, with its capability to clinically examine the debris left from 
the target, this collider should enable a far deeper understanding of the origin 
and structure of the non-perturbative sea than we have ever been 
able to imagine before. 

The generalized parton distributions, which will be thoroughly 
explored at 12 GeV 
and then exploited as source of information about the distribution of angular 
momentum on the valence quarks~\cite{Ji:1996ek,Thomas:2008ga},
will serve as a vital source of 
information on the orbital angular momentum carried by sea quarks. It 
remains to be seen whether they can also be used to investigate the gluon 
angular momentum but if a method were to be found it would be very valuable 
indeed. The understanding of the potential for 
transversity to yield information 
on the distribution of orbital angular momentum within the proton is at an 
earlier stage but there are clear indications of its 
potential which needs to be 
developed further.

As a tool to investigate the quark and gluon structure of atomic nuclei, a 
fundamental issue for modern nuclear physics, the  MEIC offers some 
remarkable new possibilities. The study of the iso-vector EMC 
effect~\cite{Cloet:2009qs} could be dramatically advanced if one could 
make a comparison of the $(e^-,\bar{\nu}_e)$ and $(e^+,\nu_e)$ reactions 
on a variety of heavy nuclei. The ability to identify fragments of the final 
nucleus offers potentially novel ways to test 
explanations  of the EMC effect. We 
expect the beam quality to be such that this machine should also be suitable 
for investigations involving parity violation, 
which also offers a novel look inside 
hadron and nucleon structure.

A number of groups have just begun to explore the 
potential of this machine for 
studying charmed systems -- from the production on free 
nucleons, to in-medium 
modification to their role as a 
possible tool to determine the gluon angular 
momentum. This is an extremely promising area 
that merits detailed examination.

Finally, one cannot imagine constructing a machine such as MEIC, with its 
high level of polarization and luminosity, 
as well as its expanded range of invariant 
mass, without exploring its potential for precision tests of 
the Standard Model. For 
the present there is nothing 
to report in this area, but I would certainly encourage 
some of the younger members of the community who may actually live long 
enough to finish such an experiment, to begin to contemplate the possibilities.

\section{Concluding Remarks}
It is clear that one or more electron-ion 
colliders may well have an important role 
to play in the development of nuclear physics over the next few 
decades~\cite{Roadmaps}. The MEIC, which is under intense study at 
Jefferson Lab, represents a cost effective first stage towards the 
very high energy ELIC. Nevertheless, MEIC appears to have a very 
impressive physics program associated with it. We certainly welcome all 
members of the DIS community who are excited by its potential and would 
like to contribute.

\section{Acknowledgments}
It is a pleasure to acknowledge the tremendous amount of effort that has so far 
been devoted to this important project by a number of staff and users at 
Jefferson Lab. I would particularly like to thank 
S.~Bogacz, P.~Chevtsov, Ya.~Derbenev, R.~Ent, G.~Krafft, T.~Horn, A.~Hutton, 
C.~Hyde-Wright, R.~Li, B.~Yunn, Y.~Zhang, F.~Klein and P.~Nadel-Turonski and 
C. Weiss, without whom this presentation would not have been possible.
This work was supported by the the U.S. Department of Energy under 
Contract No. DE-AC05-06OR23177, under which Jefferson Science Associates,
LLC operates Jefferson Laboratory. 

\section{Bibliography}
%
%
%
 %
%
\begin{footnotesize}



%

\end{footnotesize}


\end{document}